\documentclass[9pt,conference,a4paper]{IEEEtran}
\usepackage{times}

\usepackage[english]{babel}
\usepackage[T1]{fontenc}
\usepackage[utf8]{inputenc}

\usepackage{amsmath,amssymb}
\usepackage{xcolor}
\usepackage{colortbl}
\usepackage{booktabs}
\usepackage{graphicx}
\usepackage{tabularx}

\usepackage{pgfplots}
\pgfplotsset{compat=newest}
\pgfplotsset{plot coordinates/math parser=false}
\newlength\figureheight
\newlength\figurewidth 

\graphicspath{{images/}}

\title{Reconstruction of Videos Taken by a \\ Non-Regular Sampling Sensor}
\author{%
{Markus Jonscher, Jürgen Seiler, Michel Bätz, Thomas Richter, Wolfgang Schnurrer, and André Kaup}
\vspace{1.6mm}\\
\fontsize{10}{10}\selectfont\itshape
Multimedia Communications and Signal Processing \\
Friedrich-Alexander University Erlangen-Nürnberg (FAU), Cauerstr. 7, 91058 Erlangen, Germany \\
\,\\ 
\fontsize{9}{9}\selectfont\ttfamily\upshape
markus.jonscher@FAU.de
\vspace*{-0.25cm}

\vspace{1.2mm}\\
\fontsize{10}{10}\selectfont\rmfamily\itshape

\fontsize{9}{9}\selectfont\ttfamily\upshape
}
\begin{document}
\maketitle

\begin{figure}[b]
\parbox{\hsize}{\em

000-0-0000-0000-0/00/\$31.00 \ \copyright 2015 IEEE
}\end{figure}

\begin{abstract}
Recently, it has been shown that a high resolution image can be obtained without the usage of a high resolution sensor. The main idea has been that a low resolution sensor is covered with a non-regular sampling mask followed by a reconstruction of the incomplete high resolution image captured this way.
In this paper, a multi-frame reconstruction approach is proposed where a video is taken by a non-regular sampling sensor and fully reconstructed afterwards. By utilizing the temporal correlation between neighboring frames, the reconstruction quality can be further enhanced. Compared to a \mbox{state-of-the-art} single-frame reconstruction approach, this leads to a visually noticeable gain in PSNR of up to $\mathbf{1.19}$~dB on average.
\\[1\baselineskip]
\end{abstract}

\begin{keywords}
Video Processing, Non-Regular Sampling, Temporal Inpainting, Signal Extrapolation, Resolution Enhancement
\end{keywords}

\section{Introduction}
\label{sec:intro}
\begin{figure}[t]
	\centering
	\def\svgwidth{\columnwidth}	
	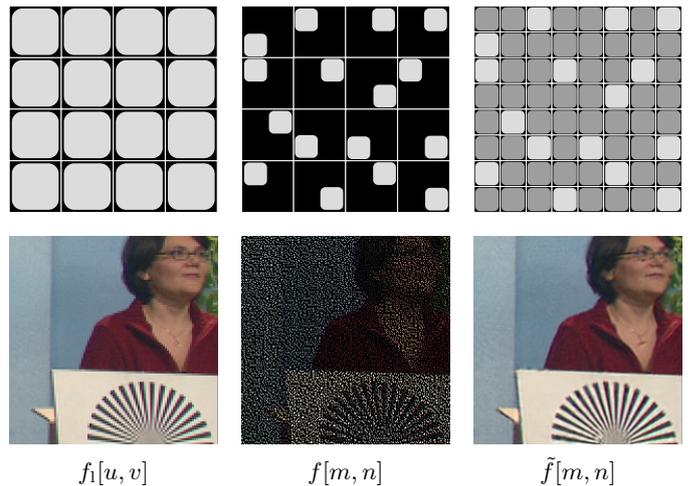
	\caption{Left: Low resolution sensor. Middle: Masked low resolution sensor. Right: Reconstructed high resolution image.}
	\label{fig:reconstruction_pipeline}
\end{figure}

In the majority of all multimedia applications, high resolution (HR) images or videos are required. The best way to achieve this is by using a camera sensor that gives the preferred quality right away. In some cases, such as high costs for multiple HR sensors in multi-view scenarios or high energy consumption of such sensors in mobile devices, it is required to use low resolution (LR) sensors. An HR output, however, is still desirable. One possibility to achieve this may be single-image super-resolution where a reasonable HR image can be obtained from a single LR image \cite{Zhang2012, Peleg2014}. Another possibility to get an HR image has been shown in~\cite{Schoeberl2011a} where an LR sensor is covered with a non-regular sampling mask. The incomplete HR image that is captured this way has to be reconstructed afterwards.
The principle of this approach is shown in Figure~\ref{fig:reconstruction_pipeline}.
On the left, an LR sensor that gives the LR image $f_{\mathrm{l}}[u,v]$ is displayed. The area of the sensor that is sensitive to light is denoted in light-gray and $(u,v)$ depict the spatial coordinates on the LR grid. 
In the middle, an LR sensor that is covered with a non-regular sampling mask can be seen. Each large pixel of this LR sensor is divided into four quadrants, where three of them are randomly covered. As a consequence, only $1/4$ of the large pixel is sensitive to light anymore. This leads to the HR image $f[m,n]$ which is only partially available due to the masking. $(m,n)$ depict the spatial coordinates on this HR grid.
Since the per-pixel area sensitive to light of such a covered LR sensor is equal to that of an HR sensor, the noise behavior is similar.
As mentioned in~\cite{Schoeberl2011a}, Frequency Selective Extrapolation (FSE) is used for the reconstruction of these missing pixels. This leads to the full HR image $\tilde{f}[m,n]$ on the right where reconstructed pixels are denoted in dark-gray.

In this paper, a multi-frame reconstruction approach is proposed where a video is captured by a non-regular sampling sensor. The temporal correlation between neighboring frames is then exploited to enhance the overall reconstruction quality.
This is achieved by projecting pixel information from neighboring frames into the area of missing pixels of the current frame to be extrapolated. This way, new sampling points are defined which are useful for reconstruction, since the performance of FSE highly depends on the number of available sampling points. For the projection of pixels from one frame into another frame, motion vectors are required. They are computed by using a modified block-based motion estimation algorithm~\cite{Santamaria2012}. 

The paper is organized as follows: The next section introduces a \mbox{state-of-the-art} method to reconstruct incomplete images and Section~\ref{sec:proposed_setup} presents the proposed multi-frame reconstruction approach for non-regular sampled video data. Simulation results are given in Section~\ref{sec:results} and Section~\ref{sec:conclusion} concludes this contribution.

\section{State-of-the-art single-frame reconstruction}
\label{sec:fse}
\begin{figure}[t]
	\centering
	\def\svgwidth{\columnwidth}	
	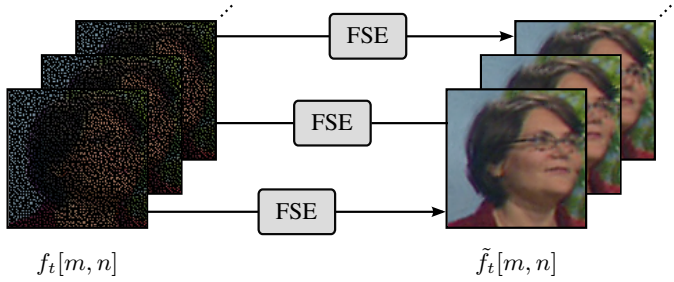
	\caption{State-of-the-art single-frame approach (FSE-SF) for videos.}
	\label{fig:single_frame_approach}
\end{figure}

In Figure~\ref{fig:single_frame_approach}, a scene is captured by a non-regular sampling sensor giving a sampled video which consists of frames $f_t[m,n]$. 
The fully reconstructed frames $\tilde{f}_t[m,n]$ are then obtained by applying Frequency Selective Extrapolation (FSE) independently on all frames $f_t[m,n]$. Up to this step, this corresponds to the reconstruction of single frames~\cite{Schoeberl2011a} and is from now on referred to as FSE-SF. FSE iteratively generates the sparse signal model
\begin{equation}
	g[m,n] = \sum_{(k,l)\in\mathcal{K}}\hat{c}_{(k,l)}\varphi_{(k,l)}[m,n]
\end{equation}
as a superposition of Fourier basis functions $\varphi_{(k,l)}[m,n]$ weighted by the expansion coefficients $\hat{c}_{(k,l)}$. This is done in a block-wise manner for each frame of the video.
The set $\mathcal{K}$ contains the indices of all basis functions that have been selected for model generation. In every iteration, one basis function gets selected and before being added to the model, its corresponding weight is estimated. The model $g[m,n]$ is then used to replace missing pixels in the corresponding HR frame $f[m,n]$. In this paper, FSE additionally uses an optimized processing order~\cite{Seiler2011}. 
Alternatively, it is also possible to employ other algorithms like Natural Neighbor Interpolation \cite{Sibson1981} or Steering Kernel Regression \cite{Takeda2007} for reconstruction. Since it has been shown in \cite{Schoeberl2011a} that FSE outperforms other methods, only FSE is regarded in this paper.
\begin{figure}[t]
	\centering
	\def\svgwidth{\columnwidth}	
	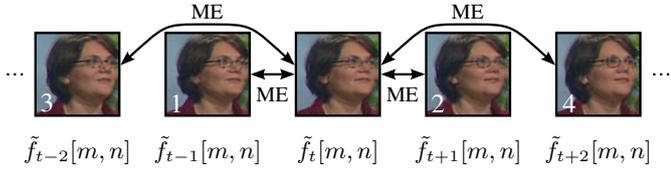
	\caption{Motion estimation (ME) between the current frame $\tilde{f}_t[m,n]$ to be extrapolated and a certain number of neighboring frames.}
	\label{fig:prerec}
\end{figure}

Recently, it has been shown in \cite{Jonscher2014} and \cite{Richter2014} that the reconstruction quality for images captured by non-regular sampling sensors in multi-view scenarios can be further enhanced by utilizing the spatial correlation between neighboring views.
In this scenario where a video is captured by a single non-regular sampling sensor it is also of interest to utilize dependencies between neighboring frames. Therefore, temporal correlation may be exploited as it is done for instance in super-resolution~\cite{Park2003} or video coding~\cite{Sullivan2012}.

\section{Proposed multi-frame reconstruction}
\label{sec:proposed_setup}
Since the performance of FSE highly depends on the number of available sampling points, it is of interest to use pixel information from neighboring frames as additional support for the reconstruction process. This approach is called multi-frame FSE (\mbox{FSE-MF$_N$}), where $N$ is the number of utilized support frames for the extrapolation of one frame. To exploit temporal correlation between neighboring frames, motion estimation (ME) is required. ME should not be applied directly on the sampled frames $f_t[m,n]$, since the large area of missing pixels leads to impractical results. Therefore, an initial reconstruction of these frames is necessary and it is advisable to use an effective algorithm such as FSE to get good results from ME. 

As shown in Figure~\ref{fig:prerec}, ME is applied between each frame $\tilde{f}_t[m,n]$ of the previously reconstructed HR video and a certain number of neighboring support frames and vice versa. The order of applying these different MEs can also be seen in this figure. It starts with an ME between $\tilde{f}_t[m,n]$ and the adjacent frame $\tilde{f}_{t-1}[m,n]$. The second ME is performed between $\tilde{f}_t[m,n]$ and $\tilde{f}_{t+1}[m,n]$, followed by $\tilde{f}_{t-2}[m,n]$, $\tilde{f}_{t+2}[m,n]$, and so on.

In this paper, a block-based ME algorithm is employed which is exemplarily described for the first ME. Instead of performing a straightforward block-matching as in \cite{Santamaria2012}, a modified matching is developed. Since it is desired to only use reliable pixels and not already extrapolated pixels, only original samples are considered for matching. Due to the knowledge of the sampling mask, the position of these pixels is available. A small square window $W$ is then taken around these pixels in $\tilde{f}_{t-1}[m,n]$. $W$ includes one original sample in the center surrounded by reconstructed and few other original samples. For each of these reference blocks, the best matching block in $\tilde{f}_t[m,n]$ is searched within a certain search range. To determine the best match between two blocks, sum of absolute differences (SAD) is employed. The relative position between these two blocks is represented by a motion vector (MV) computed with integer precision. This MV can be used to project pixels from one frame into a neighboring frame.
\begin{figure}[t]
	\centering
	\def\svgwidth{\columnwidth}	
	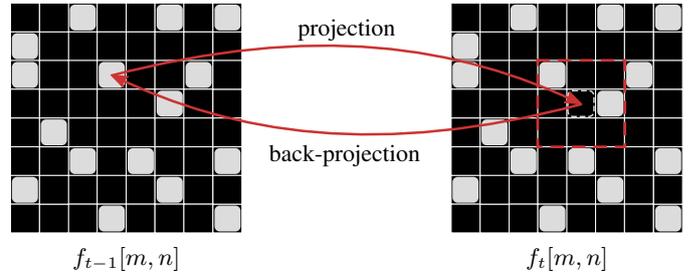
	\caption{Consistency check for refining the motion vectors.}
	\label{fig:cross_check}
\end{figure}
\begin{figure}[]
	\centering
	\def\svgwidth{\columnwidth}	
	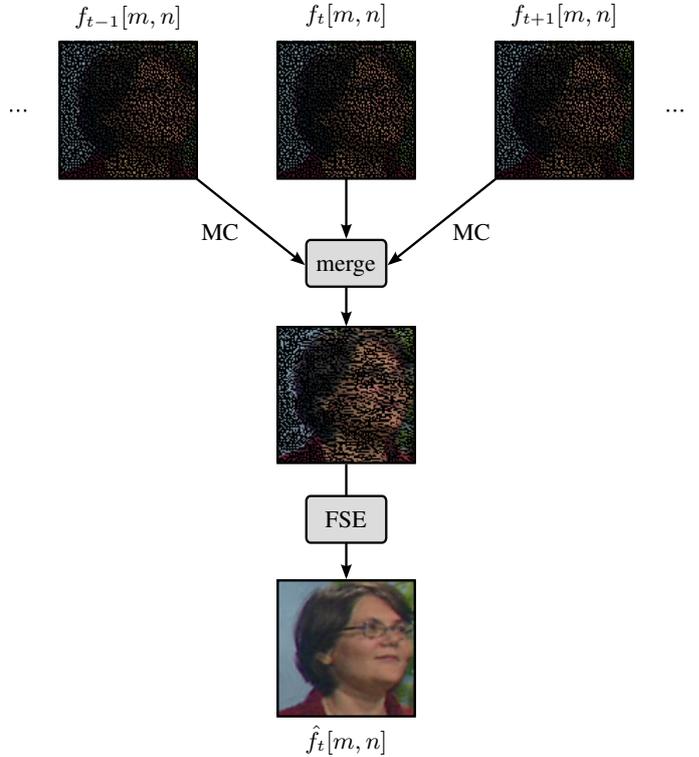
	\caption{Proposed multi-frame approach (FSE-MF$_N$) exemplarily shown for one specific frame and two support frames.}
	\label{fig:multi_frame_approach}
\end{figure}

After MV computation, a consistency check as shown in Figure~\ref{fig:cross_check} is performed. There, an original sample of $f_{t-1}[m,n]$ is projected into a neighboring frame $f_t[m,n]$. When it points onto the position of another original sample, the corresponding MV is removed, since there is already pixel information available. If it points onto a position between original samples, which is desirable, the median of all MVs that lie in a $3\times 3$ neighborhood (denoted by the red dashed square) is chosen and projected back into $f_{t-1}[m,n]$. This kind of processing is necessary, since on these positions no MVs are available. When the back-projection does not fall onto the same pixel where the projection started, the corresponding MV is also removed, since this pixel is probably projected onto a wrong position.

After all MVs are refined by the consistency check, these vectors can be used for motion compensation (MC). This means that reliable pixel information from neighboring support frames can be projected into the current block to be extrapolated. 
In Figure~\ref{fig:multi_frame_approach}, this process is exemplarily shown for FSE-MF$_2$. 
After ME is applied between $\tilde{f}_t[m,n]$ and two adjacent frames $\tilde{f}_{t-1}[m,n]$ and $\tilde{f}_{t+1}[m,n]$, the corresponding MVs are refined. By means of these MVs, the sampled frames $f_{t-1}[m,n]$ and $f_{t+1}[m,n]$ from the neighborhood are projected onto $f_{t}[m,n]$. Missing pixels in $f_{t}[m,n]$ are then replaced with pixels of the projected frames in the order specified in Figure~\ref{fig:prerec}. The result is a sampled frame which includes an extended set of sampling points. Now, FSE has less pixels to extrapolate and more pixel information that can be used during model generation. The final reconstruction by FSE leads to $\hat{f}_{t}[m,n]$. To obtain the full HR video, these steps have to be repeated for each frame $f_t[m,n]$.

\section{Simulation Results}
\label{sec:results}
\begin{figure}[t]
	\centering
	\def\svgwidth{\columnwidth}	
	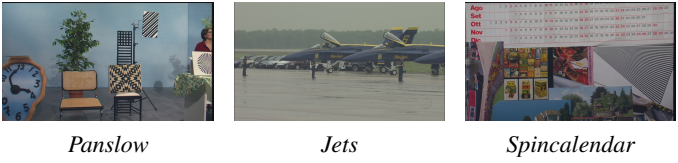
	\vspace*{-0.6cm}
	\caption{Three different $720$p test sequences used for simulation.}
	\label{fig:test_sequences}
\end{figure}

The proposed multi-frame reconstruction approach is evaluated for three $720$p test sequences as shown in Figure~\ref{fig:test_sequences}, each with a different type of motion.
\begin{table}[t]
	\caption{FSE parameters used during simulation.}
	\label{tab:fse_parameter}
	\vspace*{-0.2cm}
	\centering	
	\begin{tabularx}{\columnwidth}{p{0.55cm}lc}
		\toprule
		& Block size                                        &  $4 \times 4$   \\
		& Border width                                      &      $14$       \\
		& FFT size                                          & $32 \times 32$  \\
		& Iterations                                        &     $100$       \\
		& Decay factor $\hat{\rho}$                         &     $0.7$       \\
		& Orthogonality deficiency compensation $\gamma$    &     $0.5$       \\
		& Weighting of already reconstructed areas $\delta$ &     $0.5$       \\ \bottomrule
	\end{tabularx}
\end{table}
Selecting the first $100$ frames of each sequence leads to a comprehensive test set of $300$ frames. For simulation and evaluation only the luminance is used and for generating the example images, FSE is applied on the three RGB color channels independently. In the \emph{Panslow} sequence, there is almost no motion at first and around frame $40$, a translational motion in horizontal direction starts. The \emph{Jets} sequence consists of a global zoom and minor motion within the scene. The \emph{Spincalendar} sequence contains only rotational motion. The same sampling mask is used for every frame of each sequence and all $100$ frames are then reconstructed using both the proposed multi-frame reconstruction approach (FSE-MF) and the \mbox{state-of-the-art} single-frame approach (FSE-SF). All relevant parameters for FSE that are used during simulation are listed in Table~\ref{tab:fse_parameter}. For an extensive discussion of these parameters please refer to \cite{Schoeberl2011a}. All PSNR results are then computed excluding a margin of 4 pixels to avoid the influence of artifacts due to the black border of some frames.
\begin{figure}[t]
	\centering
%
%
\definecolor{mycolor1}{rgb}{0.00000,0.49804,0.00000}%
\begin{tikzpicture}

\begin{axis}[%
width=0.85\columnwidth,
height=0.555205\columnwidth,
at={(0\columnwidth,0\columnwidth)},
scale only axis,
xmin=0,
xmax=9,
xlabel={Number of support frames},
xmajorgrids,
ymin=0,
ymax=1.4,
ylabel={Average PSNR gain [dB]},
ymajorgrids,
legend style={at={(0.03,0.97)},anchor=north west,legend cell align=left,align=left,draw=black}
]
\addplot [color=blue,solid,mark size=1.6pt,mark=*,mark options={solid,fill=white}]
  table[row sep=crcr]{%
1	0.282433804390525\\
2	0.459950676096332\\
3	0.585698526411398\\
4	0.714215959974197\\
5	0.818409612776946\\
6	0.915854711905897\\
7	0.988326903335602\\
8	1.05957168384308\\
};
\addlegendentry{Panslow};

\addplot [color=red,solid,mark size=1.4pt,mark=square*,mark options={solid,fill=white}]
  table[row sep=crcr]{%
1	0.33423376243012\\
2	0.508502468696086\\
3	0.623962768835367\\
4	0.692489482503801\\
5	0.738812194221087\\
6	0.76958619620188\\
7	0.781698595978454\\
8	0.785778454078566\\
};
\addlegendentry{Jets};

\addplot [color=mycolor1,solid,mark size=1.9pt,mark=triangle*,mark options={solid,rotate=180,fill=white}]
  table[row sep=crcr]{%
1	0.319778419310182\\
2	0.526002892943426\\
3	0.688716535605\\
4	0.815273474371441\\
5	0.948105605828114\\
6	1.04793921518615\\
7	1.13073778838673\\
8	1.1918553033804\\
};
\addlegendentry{Spincalendar};

\end{axis}
\end{tikzpicture}%
	\caption{Comparison of FSE-MF against FSE-SF showing the average gain in PSNR for all three test sequences over the number of utilized support frames for FSE-MF.}
	\label{fig:plot_results_support_frames}
\end{figure}
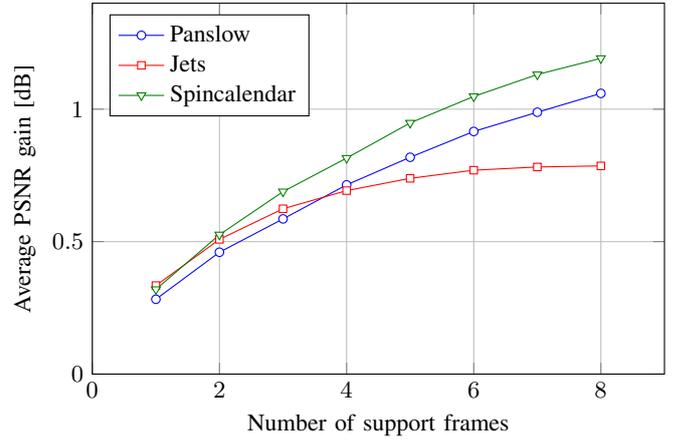
\begin{figure}[t]
	\centering
%
%
\definecolor{mycolor1}{rgb}{0.00000,0.49804,0.00000}%
\begin{tikzpicture}

\begin{axis}[%
width=0.85\columnwidth,
height=0.56085\columnwidth,
at={(0\columnwidth,0\columnwidth)},
scale only axis,
xmin=-2,
xmax=102,
xlabel={Frame number},
xmajorgrids,
ymin=-0.1,
ymax=1.1,
ylabel={PSNR gain [dB]},
ymajorgrids,
legend style={at={(0.03,0.97)},anchor=north west,legend cell align=left,align=left,draw=black}
]
\addplot [color=blue,solid,mark size=1.4pt,mark=*,mark options={solid,fill=white}]
  table[row sep=crcr]{%
1	-0.00346290264582194\\
2	0.00430034573239624\\
3	-0.00991163954592267\\
4	-0.00649900048190588\\
5	7.43703750316627e-06\\
6	-0.00166567417497632\\
7	0.000328818583689383\\
8	0.00412318266090139\\
9	-0.0047468850489274\\
10	-0.00301531621915174\\
11	0.00196954740317778\\
12	-0.000452866765868976\\
13	0.00885629415449074\\
14	0.0194500488718283\\
15	0.0209079526419096\\
16	0.0295226872638601\\
17	0.0365568502625457\\
18	0.0387508121077609\\
19	0.0561694825996604\\
20	0.0533970547572267\\
21	0.0671161658155555\\
22	0.0571248630878891\\
23	0.0757295519595189\\
24	0.0742685067898101\\
25	0.0858969847473467\\
26	0.0783847714022663\\
27	0.0756501558444569\\
28	0.0785806331832006\\
29	0.0871474684924998\\
30	0.0737233151054077\\
31	0.089155983141687\\
32	0.0915998519954044\\
33	0.0918242462636059\\
34	0.0902229706214683\\
35	0.0997796449665991\\
36	0.0972236468420142\\
37	0.0823165022087515\\
38	0.0770117156511922\\
39	0.104351206029829\\
40	0.0923725788861418\\
41	0.0841949036570355\\
42	0.0825722654853038\\
43	0.140478937051519\\
44	0.167833962591157\\
45	0.187442607464952\\
46	0.181912211538709\\
47	0.209318413676691\\
48	0.188838607704444\\
49	0.277863480937341\\
50	0.289654191172168\\
51	0.308222705750079\\
52	0.356056371731338\\
53	0.413843765822971\\
54	0.400572098246826\\
55	0.450122241051979\\
56	0.493294384326784\\
57	0.632819814071606\\
58	0.608532487600552\\
59	0.615016773478089\\
60	0.621178554803389\\
61	0.671009891764797\\
62	0.74195917549671\\
63	0.796085239442021\\
64	0.770559317108166\\
65	0.806294042707922\\
66	0.896317921934347\\
67	0.938299678397328\\
68	0.912175888768509\\
69	0.985082577295373\\
70	0.947918294826462\\
71	0.989581618831309\\
72	0.948950017067034\\
73	0.948474565561934\\
74	0.949282700551041\\
75	0.988638399808806\\
76	1.04257766082237\\
77	1.00531815022511\\
78	0.995819468132474\\
79	1.04066379667412\\
80	0.974705287968529\\
81	1.03312204570999\\
82	0.884199829947818\\
83	0.984625333040107\\
84	1.0046498988269\\
85	0.977453329588172\\
86	1.04191340938757\\
87	1.00931011651178\\
88	1.00994725664775\\
89	1.00776445562448\\
90	0.980357041533274\\
91	0.968637258241074\\
92	0.962848167537384\\
93	0.927332125248014\\
94	0.935994441072161\\
95	0.953468449215926\\
96	0.978293178493352\\
97	0.93885671003606\\
98	0.922246938793592\\
99	0.930953764676424\\
100	0.569544395725064\\
};
\addlegendentry{Panslow};

\addplot [color=red,solid,mark size=1.1pt,mark=square*,mark options={solid,fill=white}]
  table[row sep=crcr]{%
1	0.295397991100138\\
2	0.489331223103036\\
3	0.544647772804737\\
4	0.567834554046811\\
5	0.506583999520473\\
6	0.552305405221162\\
7	0.576462092349587\\
8	0.579697975783347\\
9	0.56960368602067\\
10	0.527715473141761\\
11	0.525970794964039\\
12	0.583525033226941\\
13	0.509934919234787\\
14	0.520248268532796\\
15	0.514846633149112\\
16	0.521436684579328\\
17	0.547176535764656\\
18	0.577412689997175\\
19	0.562647432006202\\
20	0.548782731061074\\
21	0.583837214947827\\
22	0.518237034517995\\
23	0.530639504919343\\
24	0.537818109724938\\
25	0.544870026891317\\
26	0.513571177194549\\
27	0.542411688120133\\
28	0.54706960043697\\
29	0.539611744856941\\
30	0.512792657975453\\
31	0.544916266183563\\
32	0.548247707458337\\
33	0.495918591122006\\
34	0.510327706368436\\
35	0.592264065004422\\
36	0.474296927425883\\
37	0.487889595385489\\
38	0.59218127699215\\
39	0.504048284069221\\
40	0.50962595305883\\
41	0.511195478844662\\
42	0.487692274364989\\
43	0.510274578275904\\
44	0.485989639005787\\
45	0.489838299443335\\
46	0.481674089983365\\
47	0.445057185506776\\
48	0.493315568935422\\
49	0.513826142660385\\
50	0.489745416308409\\
51	0.542505133376174\\
52	0.457062968690558\\
53	0.486180521646389\\
54	0.520873031332549\\
55	0.484142624706664\\
56	0.548655170939618\\
57	0.486970773007172\\
58	0.54239311791504\\
59	0.513419304263941\\
60	0.551517211268923\\
61	0.550520072065126\\
62	0.470581568050477\\
63	0.503589010517047\\
64	0.514062771350666\\
65	0.46732728901452\\
66	0.467426319822273\\
67	0.514227934464024\\
68	0.52253061126919\\
69	0.486569756949564\\
70	0.469221989306462\\
71	0.523242218007297\\
72	0.472635048586227\\
73	0.48768965264783\\
74	0.491518968906618\\
75	0.495328448226644\\
76	0.487193168443397\\
77	0.447424797358508\\
78	0.49631740352968\\
79	0.523047150151385\\
80	0.498460648930468\\
81	0.494775736132688\\
82	0.501550664837175\\
83	0.426927319368545\\
84	0.449624388203127\\
85	0.433135185273919\\
86	0.477902836014067\\
87	0.456820410886898\\
88	0.506402635129788\\
89	0.484870851619043\\
90	0.515435431381164\\
91	0.492596800775544\\
92	0.483310227787051\\
93	0.532369402991414\\
94	0.504722099755632\\
95	0.471225263455857\\
96	0.470702345153811\\
97	0.538261371549076\\
98	0.566390349990257\\
99	0.528373136073995\\
100	0.331470026894401\\
};
\addlegendentry{Jets};

\addplot [color=mycolor1,solid,mark size=1.5pt,mark=triangle*,mark options={solid,rotate=180,fill=white}]
  table[row sep=crcr]{%
1	0.292262955511145\\
2	0.428109306255795\\
3	0.444289881988304\\
4	0.448370969027241\\
5	0.471912671383503\\
6	0.457379891691971\\
7	0.425647952800922\\
8	0.407350125911687\\
9	0.417849633471072\\
10	0.424448453931596\\
11	0.416670210047023\\
12	0.459584725891553\\
13	0.431944140210792\\
14	0.435965050003386\\
15	0.457539675020673\\
16	0.444316467781487\\
17	0.449517728252687\\
18	0.473633763095581\\
19	0.436321543916552\\
20	0.456337570943322\\
21	0.46167724252571\\
22	0.468611639466296\\
23	0.496745033938844\\
24	0.497173605086196\\
25	0.458807991414204\\
26	0.535615194448162\\
27	0.464621896819022\\
28	0.492373397355067\\
29	0.516203146169399\\
30	0.512847398710512\\
31	0.492606980828995\\
32	0.496859301062145\\
33	0.527711927715792\\
34	0.52756705270297\\
35	0.507183635966243\\
36	0.520319585651851\\
37	0.550583226890691\\
38	0.540317016376608\\
39	0.509307445264923\\
40	0.519451071946179\\
41	0.530612549546131\\
42	0.536263189318639\\
43	0.537191022417989\\
44	0.533211752454317\\
45	0.535078924439528\\
46	0.579060147840906\\
47	0.589946811219182\\
48	0.550022933067961\\
49	0.563246709347187\\
50	0.548009932137305\\
51	0.541889490634382\\
52	0.566606199511597\\
53	0.536796658987576\\
54	0.558342623567384\\
55	0.576186728666105\\
56	0.5555678236117\\
57	0.53734258606821\\
58	0.567288907747042\\
59	0.534053946760213\\
60	0.554020557716829\\
61	0.573891594802475\\
62	0.581276238310519\\
63	0.531272201192735\\
64	0.551971718718711\\
65	0.567395954717608\\
66	0.567048118620619\\
67	0.54241834240765\\
68	0.56831885006012\\
69	0.584686697535833\\
70	0.562200738708437\\
71	0.579892275092313\\
72	0.556308979445852\\
73	0.579542162861809\\
74	0.572183705987527\\
75	0.566467110174983\\
76	0.586192495026737\\
77	0.59002723479156\\
78	0.566720644155708\\
79	0.584870192573021\\
80	0.559511727665353\\
81	0.575019522540114\\
82	0.585956289599864\\
83	0.57367368326004\\
84	0.589530153650571\\
85	0.568712731404862\\
86	0.551788694386943\\
87	0.57711529955705\\
88	0.567016298273657\\
89	0.564635651557452\\
90	0.60491695975125\\
91	0.60460023415429\\
92	0.581302861992931\\
93	0.574899980325249\\
94	0.577149394431338\\
95	0.586541488310644\\
96	0.593134761297041\\
97	0.587248915461696\\
98	0.606800033153789\\
99	0.587665907761316\\
100	0.363607448086604\\
};
\addlegendentry{Spincalendar};

\end{axis}
\end{tikzpicture}%
	\caption{Gain in PSNR for FSE-MF$_2$ compared to FSE-SF over all frames of each test sequence.}
	\label{fig:plot_results_all_frames}
\end{figure}
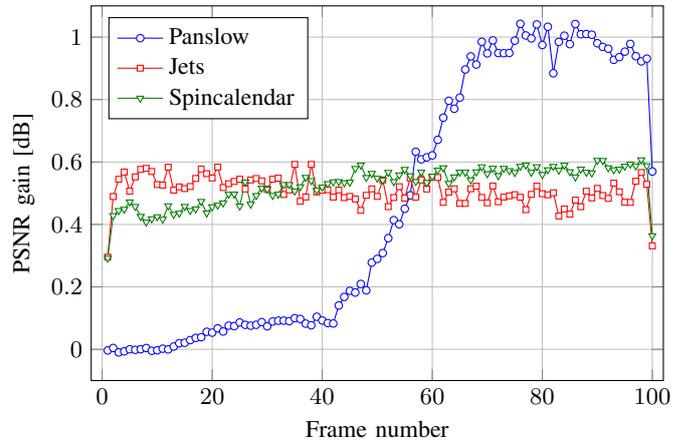

In Figure~\ref{fig:plot_results_support_frames}, the average gain in PSNR for FSE-MF over \mbox{FSE-SF} is plotted over the number of utilized support frames (\mbox{FSE-MF$_{1-8}$}). 
It can be seen that even for FSE-MF$_1$, an average gain of up to $0.33$~dB is achievable. It is also noticeable that the more support frames are employed, the higher the overall gain gets. For FSE-MF$_8$, a maximum average gain of $1.19$~dB can be achieved for \emph{Spincalendar}.
Since for half of the \emph{Panslow} sequence there is only small motion and therefore only small gains, \emph{Panslow} lies below \emph{Spincalender}. Motion estimation in \emph{Jets} is more difficult especially for a larger number of support frames and therefore, the average gains for this sequence are not as high as for the other two sequences.

\begin{figure*}[ht]
	\centering
	\def\svgwidth{\textwidth}
	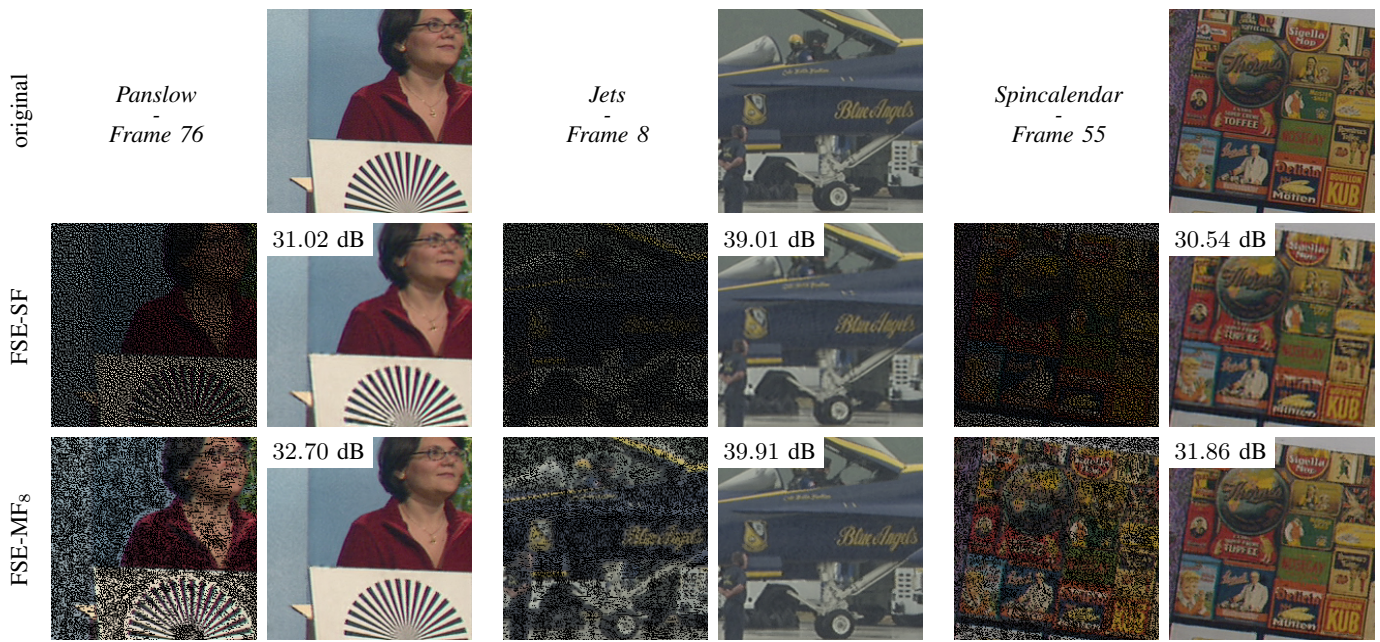
	\caption{Three image detail examples comparing the \mbox{state-of-the-art} FSE-SF and the proposed FSE-MF using eight support frames.}
	\label{fig:image_examples}
\end{figure*}

Figure~\ref{fig:plot_results_all_frames} shows the gain for FSE-MF$_2$ over FSE-SF, plotted for each frame of the three test sequences. It can be seen that when there is only little motion as in the beginning of \emph{Panslow}, the gain from using support frames is very small. For sequences with constant motion such as \emph{Jets} or \emph{Spincalendar} there is also a constant gain over all frames. It is observable that the utilized ME works best for translational motion, however, there are also reasonable results achievable for other types of motions. Additionally, it can be seen that for the first and the last frame of each sequence the gain decreases. Since in this example FSE-MF$_2$ uses two adjacent support frames, frame $1$ and frame $100$ can only exploit one support frame which corresponds to FSE-MF$_1$. This applies analogously to \mbox{FSE-MF$_{3-8}$}.

A maximum gain of $1.76$~dB can be reached for frame $79$ of the \emph{Panslow} sequence using FSE-MF$_8$. Occasionally, the gain may be negative near the first or the last frame and when there is almost no motion. In the worst case, values of up to $-0.1$~dB are reached. Since this occurs very rarely and is visually not noticeable, this effect needs no special treatment.

Figure~\ref{fig:image_examples} shows three image detail examples for the comparison of FSE-SF and the proposed FSE-MF where for the latter the number of support frames is chosen to be $8$. For visual comparison, the original frames can be seen in the upper row. The middle row shows how FSE-SF is able to reconstruct the corresponding non-regular sampled frames. The last row shows the visual results for \mbox{FSE-MF$_8$}. It can clearly be seen that more pixel information is available in the corresponding sampled frames which leads to an enhanced reconstruction quality. Using only FSE-SF, it is noticeable that for all frames fine details cannot be reconstructed and ringing artifacts or blurring occurs. With the proposed FSE-MF, all frames appear sharper, fine details are reconstructed and especially small text is better readable.

\vspace*{0.3cm}
\section{Conclusion}
\label{sec:conclusion}
In this paper, a multi-frame reconstruction approach has been proposed that is able to reconstruct a video captured by a non-regular sampling sensor. An enhanced reconstruction quality could be achieved by using a modified block-based motion estimation algorithm and by utilizing pixel information from neighboring frames. Compared to a \mbox{state-of-the-art} single-frame reconstruction approach, this led to a visually noticeable gain in PSNR of up to $1.76$~dB for a specific frame and an average gain of up to $1.19$~dB.

For future research, the influence of different image reconstruction algorithms during the initial reconstruction step and other motion estimation algorithms such as various block-based motion estimations~\cite{Santamaria2012} or optical flow methods \cite{Sanchez2013} will be investigated. 
To enhance the reconstruction quality when there is no motion, varying the non-regular sampling mask over time is another topic for future research.

\section*{Acknowledgement}
This work has been supported by the Deutsche Forschungsgemeinschaft (DFG) under contract number KA 926/5-1.

\bibliographystyle{IEEEtran}
\bibliography{literature}

\begin{thebibliography}{10}
\providecommand{\url}[1]{#1}
\csname url@samestyle\endcsname
\providecommand{\newblock}{\relax}
\providecommand{\bibinfo}[2]{#2}
\providecommand{\BIBentrySTDinterwordspacing}{\spaceskip=0pt\relax}
\providecommand{\BIBentryALTinterwordstretchfactor}{4}
\providecommand{\BIBentryALTinterwordspacing}{\spaceskip=\fontdimen2\font plus
\BIBentryALTinterwordstretchfactor\fontdimen3\font minus
  \fontdimen4\font\relax}
\providecommand{\BIBforeignlanguage}[2]{{%
\expandafter\ifx\csname l@#1\endcsname\relax
\typeout{** WARNING: IEEEtran.bst: No hyphenation pattern has been}%
\typeout{** loaded for the language `#1'. Using the pattern for}%
\typeout{** the default language instead.}%
\else
\language=\csname l@#1\endcsname
\fi
#2}}
\providecommand{\BIBdecl}{\relax}
\BIBdecl

\bibitem{Zhang2012}
K.~Zhang, X.~Gao, D.~Tao, and X.~Li, ``{S}ingle {I}mage {S}uper-{R}esolution
  {W}ith {N}on-{L}ocal {M}eans and {S}teering {K}ernel {R}egression,'' vol.~21,
  no.~11, pp. 4544--4556, Nov. 2012.

\bibitem{Peleg2014}
T.~Peleg and M.~Elad, ``{A} {S}tatistical {P}rediction {M}odel {B}ased on
  {S}parse {R}epresentations for {S}ingle {I}mage {S}uper-{R}esolution,''
  vol.~23, pp. 2569--2582, Jun. 2014.

\bibitem{Schoeberl2011a}
M.~Schöberl, J.~Seiler, B.~Kasper, S.~Fößel, and A.~Kaup, ``{S}parsity-based
  {D}efect {P}ixel {C}ompensation for {A}rbitrary {C}amera {R}aw {I}mages,''
  Prague, Czech Republic, May 2011, pp. 1257--1260.

\bibitem{Santamaria2012}
M.~Santamaría and M.~Trujillo, ``{A} {C}omparison of {B}lock-{M}atching
  {M}otion {E}stimation {A}lgorithms,'' in \emph{Proc. IEEE 7th Colombian
  Computing Congress}, Medellín, Colombia, October 2012, pp. 1--6.

\bibitem{Seiler2011}
J.~Seiler, \emph{{S}ignal {E}xtrapolation {U}sing {S}parse {R}epresentations
  and its {A}pplications in {V}ideo {C}ommunication}.\hskip 1em plus 0.5em
  minus 0.4em\relax Munich, Germany: Verlag Dr. Hut, December 2011.

\bibitem{Sibson1981}
R.~Sibson, ``{A} {B}rief {D}escription of {N}atural {N}eighbor
  {I}nterpolation,'' in \emph{Interpreting Multivariate Data}.\hskip 1em plus
  0.5em minus 0.4em\relax John Wiley \& Sons, 1981, pp. 21--36.

\bibitem{Takeda2007}
H.~Takeda, S.~Farsiu, and P.~Milanfar, ``{K}ernel {R}egression for {I}mage
  {P}rocessing and {R}econstruction,'' vol.~16, no.~2, pp. 349--366, Feb. 2007.

\bibitem{Jonscher2014}
M.~Jonscher, J.~Seiler, T.~Richter, M.~Bätz, and A.~Kaup, ``{R}econstruction
  of {I}mages {T}aken by a {P}air of {N}on-{R}egular {S}ampling {S}ensors
  {U}sing {C}orrelation {B}ased {M}atching,'' Paris, France, Oct. 2014, pp.
  2879--2882.

\bibitem{Richter2014}
T.~Richter, M.~Jonscher, W.~Schnurrer, J.~Seiler, and A.~Kaup,
  ``{R}econstruction of {M}ultiview {I}mages {T}aken with {N}on-{R}egular
  {S}ampling {S}ensors,'' Florence, Italy, May 2014, pp. 5789--5793.

\bibitem{Park2003}
S.~C. Park, M.~K. Park, and M.~G. Kang, ``{S}uper-{R}esolution {I}mage
  {R}econstruction: {A} {T}echnical {O}verview,'' vol.~20, no.~3, pp. 21--36,
  May 2003.

\bibitem{Sullivan2012}
G.~J. Sullivan, J.-R. Ohm, W.-J. Han, and T.~Wiegand, ``{O}verview of the
  {H}igh {E}fficiency {V}ideo {C}oding ({HEVC}) {S}tandard,'' vol.~22, pp.
  1649--1668, Sep. 2012.

\bibitem{Sanchez2013}
J.~Sánchez, E.~Meinhardt-Llopis, and G.~Facciolo, ``{TV}-{L}1 {O}ptical {F}low
  {E}stimation,'' \emph{Image Processing On Line}, vol.~3, pp. 137--150, 2013.

\end{thebibliography}

\end{document}